\documentstyle[12pt]{article}
\baselineskip=\normalbaselineskip

\begin{document}

\begin{titlepage}

\begin{flushright}
KUCP-0097 \\ 
May 1996 
\end{flushright}
\vskip 1cm

\begin{center}
{\Large  
 Higher Dimensional Self-similar Spherical Symmetric Scalar
 Field Collapse and Critical Phenomena in Black Hole Formation
}
\vskip 2cm

{\large  
Jiro Soda
}\vskip 0.7cm

{\sl  
Theoretical Physics Institute,
University of Alberta  \\
Edmonton, Alberta, Canada T6G 2J1 \\
and \\
Department of Fundamental Sciences,
FIHS,\\
Kyoto University, Kyoto 606, Japan}
\vskip 2cm
{\large  
 Kouichirou Hirata
}\vskip 0.7cm
{\sl  
Graduate School of Human and Environmental Studies,\\
Kyoto University, Kyoto 606, Japan}
\end{center}
\vskip 2cm

\begin{abstract}
 The higher dimensional spherical symmetric scalar field
 collapse problem is studied in the light of the critical
 behavior in black hole formation. To make the analysis
 tractable, the self similarity is also imposed.
  By giving a new view to the self-similar scalar
field collapse problem, we give the general formula for the
critical exponents in higher dimensions.
 In the process, the explanation of the universality of the 
critical phenomena is given within the self-similar context.
\end{abstract}
\end{titlepage}


\setlength{\normalbaselineskip}{20pt plus 0.2pt minus 01.pt}
\baselineskip=\normalbaselineskip

\section{Introduction}

 Recently, Choptuik analyzed  the spherically symmetric scalar field
collapse problem numerically and found several intriguing facts~\cite{chop}. 
Once the initial data which is characterized
by one parameter $p$  is prepared,
one can calculate the evolution of that initial data.
Then the problem is the fate of the spacetime.
If the amplitude of the field is not so large,
the imploding scalar wave will scatter away
and the flat spacetime remains.
On the contrary, if the amplitude of the  field is large enough,
the imploding scalar wave will collapse
and the apparent horizon is developed.
The final state is nothing but a black hole.
Let us consider the verge of two cases,
as the initial data is characterized by the one parameter,
there is a critical value $p^* $ which separates two regions,
the corresponding solution is now called "choptuon".
The findings by Choptuik are echoing, scaling, and universality.
The echoing is the feature of the choptuon;
which shows a discrete self-similarity
\begin{equation}
   \phi (\rho - \Delta , \tau - \Delta ) = \phi (\rho , \tau) ,
\end{equation}
where $\rho$ and $\tau$ are logarithms of proper radius $r$
and central proper time $t$.
Here $\Delta \sim 3.4$.
More interestingly, in the super-critical case  $p>p^* $,
there is the scaling law in the mass formula
if the initial data close to the critical value enough, i.e.,
\begin{equation}
 M_{B.H.} \sim ( p - p^* )^\beta,
\end{equation}
where the critical exponent $\beta$ is about $0.37$.
The universality states that the exponent $\beta$
and the echoing parameter $\Delta$ are independent
of any choice of parameter $p$ of the initial data. 

The above critical phenomena are also interesting
in the light of the cosmic censorship conjecture.
Indeed, the motivation of Choptuik lied
on the cosmic censorship conjecture.
The choptuon is characterized
by a horizonless (massless black hole) spacetime
and the occurrence of a strong curvature singularity from
the regular initial data.
The choptuon offers a new numerical counterexample
to the cosmic censorship conjecture.

It may also be related to the quantum evaporating black hole.
As is well known, when we consider the quantum effects,
the black hole is no longer black.
It emits the radiation whose strength is proportional
to the surface gravity.
Then, as the evaporation proceeds,
the rate of the emission of the energy is accelerated.
The final state might be choptuon,
and the scaling law might have some physical significance. 

Soon after his discovery,
Abrahams and Evans~\cite{AE} numerically investigated
the axially-symmetric collapse of the gravitational wave
and found that the exponent becomes $\beta=0.37$
and the echo is characterized by $\Delta=0.6$.
Furthermore, Evans and Coleman~\cite{EC} studied
the radiation fluid system numerically  and obtained $\beta=0.36$.
In this case, the choptuon shows the continuous self-similarity.
Thus, they suggested the perturbative analysis
around this self-similar solution to obtain the exponent.
Koike et. al.~\cite{koike} realized this program and 
confirmed the results of Evans and Coleman. From these results,
one can observe
that the echo strongly depends on the matter content
and hence feature of the matter.
However, it is expected that the scaling is independent
on the matter content,
i.e., there may be a universality in the strong sense.
So, we are concentrating on the scaling behavior in this paper.
Of course, there is a work by Maison~\cite{maison}
which studied the general perfect fluid
by using perturbation method and found $\beta$
depends on the equation of state.
This result should be confirmed by numerical calculation.
Chiba and one of the author (J.S.)~\cite{chiba} have also investigated
the dependence of the phenomena on the theory, more precisely,
they analyzed Brans-Dicke theory and concluded that
$\beta$ depends on the Brans-Dicke parameter
and $\Delta$ is not sensitive on that parameter.
Even if the universality in the strong sense does not exist,
the universality in the primary sense is interesting enough
to investigate and we believe it reflects the nonlinear
dynamics intrinsic to the gravity. 

Almost of the works mentioned above are numerical works,
however, it is important to give analytical models
to understand underlying physics.
At present, we know Roberts solution as such a model~\cite{roberts}.
This model shows the critical phenomena in the sense
which will be explained later.
One of the purpose of this paper is to give a new look at to this model.
Then we can understand the origin of the universality properly.

As we expect the superstring theory is the theory of everything,
it may be meaningful to consider higher dimensional collapse problem.
Moreover, from the universality point of view,
the various critical behavior should be examined.
Therefore, we shall consider the higher dimensional
scalar field collapse problem with
restriction to the spherical symmetry and the self-similarity.
If we can find the exact solution in general,
it is easy to calculate the exponents.
However, it seems difficult to obtain the exact solution in
higher dimensions. 
Thus, the aim of this paper is to give an alternative view
to critical behavior exhibited by Roberts solution
and as an application to calculate the critical exponents
of n-dimensional self-similar scalar field collapse.

The organization of this paper is as follows.
In sec.2, the Roberts solution is reviewed
and the meaning of the critical behavior is specified.
In sec.3, we present the basic equations in the appropriate form
for our purpose.
In sec.4, the origin of the universality is explained
and the general formula for the critical exponents is given.
The final section devotes to the summary of our work.

\section{Roberts solution and its critical behavior}

In this section, we review the Roberts solution~\cite{roberts}
to make the meaning of the critical phenomena precise. 
We will consider the spherical symmetric collapse of a scalar field.
Let us take the line element as
\begin{equation}
   ds^2 = - e^{2\rho} du dv + r^2 d\Omega^2 \ ,
\end{equation}
where $\rho$ and $r$ depend only on $u$ and $v$. 
Here, we take $d\Omega^2$ as a line element of 2-dimensional unit sphere.
Then, from the Lagrangian
\begin{equation}
  S =  \int d^4x \sqrt{-g} [R - {1\over 2}
      g^{\mu\nu}\partial_\mu \phi \partial_\nu \phi ]  \ ,
\end{equation}
we obtain Einstein equations:
\begin{eqnarray}
  \partial_{u} \partial_{v} r^2 + {1\over 2} e^{2\rho} &=& 0  \ , \\
 \partial_{u}^2 r - 2 \partial_{u} \rho \partial_{u} r &=& 
       - {1 \over 4} r (\partial_{u} \phi)^2  \ , \\
 \partial_{v}^2 r - 2 \partial_{v} \rho \partial_{v} r &=& 
       - {1 \over 4} r (\partial_{v} \phi)^2  \ , \\
 \partial_{u} \partial_{v} \rho - {1\over r^2} \partial_{u} r \partial_{v} r
     &=&  {1\over 4r^2} e^{2\rho} - {1 \over 4} 
    \partial_{u} \phi \partial_{v} \phi  \ , 
\end{eqnarray}
and the scalar field equation:
\begin{equation}
2\partial_{u} \partial_{v} \phi +\partial_{u} (\log r^2) \partial_{v} \phi
  + \partial_{v} (\log r^2) \partial_{u} \phi =0 \ .
\end{equation}
It should be noticed that the generalization
to n massless scalar fields does not alter the structure of the equations.
Hence, the critical exponent which we shall calculate
is the same with the case of n massless scalar fields.

Now let us impose a self-similarity on the fields such that
\begin{eqnarray}
  r &=& v f(\eta) \ ,\\
  \rho &=& \rho (\eta) \ ,\\
   \phi &=& \phi (\eta) \ ,
\end{eqnarray}
where $\eta = u/v $.
The above ansatz give equations:
\begin{eqnarray}
  ({ f f^{\prime} \over \eta})^{\prime} &=& {1 \over 4\eta^2} e^{2\rho}  \ ,\\
  f^{\prime\prime} + {2\over \eta} \rho^{\prime} [f-\eta f^{\prime}]
                &=& -{1 \over 4} f \phi^{\prime 2}  \ ,\\
  f^{\prime\prime}- 2\rho^{\prime} f^{\prime} 
            &=&  -{1 \over 4} f \phi^{\prime 2}  \ , \\
 \rho^{\prime} +\eta \rho^{\prime\prime}+
  {f^{\prime} \over f^2} (f-\eta f^\prime) &=&
 - {1\over 4f^2} e^{2\rho} - {1 \over 4} \eta \phi^{\prime 2} \ , \\
  ({ f^2 \phi^{\prime}})^{\prime} &=&  0  \ , 
\end{eqnarray}
where the prime denotes the derivative with respective to $\eta$.
 Eqs. (14) and (15) yield
\begin{equation}
  \rho^{\prime} =0  \ .
\end{equation}
It should be noticed that the solution of eqs.(13) and (18)
are independent of the number of the scalar fields.
Using eq.(18), eq.(15) becomes
\begin{equation}
  f^{\prime\prime} = - {1\over 4} f \phi^{\prime 2} \ .
\end{equation}
Eqs.(13), (18) and (19) give a closed equation system.
Eqs.(16) and (17) are redundant equations.
Without loosing generality, we can put $e^{2\rho} =1$.
Using eq.(13), we obtain
\begin{equation}
 f^2 = {1\over 4} [ \eta^2 - 2\eta + 1-p^2] \ ,
\end{equation}
where $p$ is the integration constant.
Substituting the solution (20) into eq.(19), we get
\begin{equation}
  \phi = \pm  \log {u - (1+p) v \over  u - (1-p) v } \ .
\end{equation}
Consequently, the line element becomes
\begin{equation}
 ds^2 = - du dv  + 
       {1\over 4}(u^2  -2 u v +(1-p^2) v^2 )d\Omega^2 \ .
\end{equation}

Now, we shall see how the critical phenomena arises.
It is straightforward to calculate the spacetime curvature as
\begin{equation}
     R = {p^2 u v \over 2r^4} \ .
\end{equation}
The curvature singularity occurs at $ u =(1-p) v  \  ( v >0)$ 
and $  u =(1+p) v  \ (v <0) $.
The apparent horizon is given by the equation
\begin{equation}
   \theta = {(1-p^2) v - u \over  4r^2}=0 \ ,
\end{equation}
where $\theta$ is the expansion of the outgoing null rays.
Hence, the apparent horizon is located at $ u =(1-p^2) v  $.
The structure of the spacetime depends on the parameter $p$.
In the case, $0<p<1$, the singularity becomes timelike in any region. 
There is no apparent horizon and therefore no black hole.
Physically, this case corresponds to the weak implosion of the scalar wave.
In the case, $p>1$, the singularity becomes spacelike
in the region $ v >0$, while the singularity
in the region $v < 0$ remains to be timelike.
Moreover the spacelike singularity is hidden by the apparent horizon.
Therefore, there is the critical phenomena in this system.
The critical parameter of this system is $p^* =1$.
To investigate the scaling behavior,
it is necessary to calculate the mass of the black hole
in the supercritical region. 
Due to the spherical symmetry, there is a local mass function
\begin{equation}
  M = {r\over 2}(1-g^{a b}\partial_a r \partial_b r )  \ ,
\end{equation} 
where $a,b$ runs only $v$ and $u$. 
On the apparent horizon, this mass (25) becomes
\begin{equation}
    M_{A.H.} = {p \sqrt{p^2 -1} v \over 4}  \ .
\end{equation}
Unfortunately, this quantity diverges in the asymptotic future
due to the self-similarity.
This fact is also understood from the eq.(21)
which shows the imploding scalar wave also diverges
in the asymptotic future.
In other words,
\begin{equation}
  M_{u \rightarrow - \infty} = {p^2 v \over 4}  
\end{equation}
also diverges.
Therefore, we will define the order parameter as
\begin{equation}
  {M_{A.H.} \over M_{u \rightarrow -\infty} } = {\sqrt{p^2 -1} \over p} \ .
\end{equation}
Notice that, in the present case, $p^* =1$,
hence, it shows the scaling behavior
\begin{equation}
   {M_{A.H.} \over M_{u \rightarrow -\infty} } \sim (p-p^* )^{1/2} \ ,
\end{equation}
here the exponent can be read off as $1/2$~\cite{brady1,ONT}.

In this sense, we say the self-similar spherical collapse shows the critical
behavior whose critical exponent is $1/2$.
In the following discussion on the higher dimensional critical phenomena,
this definition of the order parameter is assumed.
Of course, it is possible to make the spacetime asymptotic
flat by cutting and pasting the spacetime appropriately
as is discussed in \cite{brady1,ONT}.
As we are interested in the critical behavior,
not the cosmic censorship conjecture, we are not concerned with it anymore.

\section{Collapse dynamics as an autonomous system }

As it seems difficult to find exact solution in higher dimensions
by the method similar to that used in the previous section,
an alternative method must be found to calculate the critical exponent
in black hole formation.
For that purpose, we will use the similar formalism
used by Brady~\cite{brady2} in the slightly different context. 
Using advanced Bondi coordinates ${v,r,\theta_1, \cdots \theta_{n-2}}$,
the spherical symmetric metric can be written
\begin{equation}
  ds^2 = - g(v,r) \bar{g} (v,r) dv^2 + 2 g(v,r) dv dr
 + r^2 d\Omega^2_{n-2}  ,
\end{equation}
where $d\Omega^2_{n-2}$ is the line element of the hyper sphere
in n-dimensions.
Hence $r$ is the proper circumferential radius.
Using this parameterization, the equations of motion can be 
derived from the action (4). From $G_{rr}=8\pi T_{rr}$,
we obtain
\begin{equation}
 (\log g)_{,r} = {8\pi \over n-2} r (\phi_{,r})^2   \ .
\end{equation} 
The equation $G^r_v = 8\pi T^r_v$ leads to
\begin{equation}
 ( {\bar{g} \over g} )_{,v} = - {16\pi \over n-2}
 {r\over g} [ (\phi_{,v} )^2 + \bar{g} \phi_{,r} \phi_{,v} ] \ .
\end{equation}
Furthermore, one can find
\begin{equation}
  R_{\theta \theta} = 8\pi [ T_{\theta\theta}
 - {1\over n-2} g_{\theta\theta} ( g^{\mu\nu} T_{\mu\nu}
 + (n-2) g^{\theta\theta} T_{\theta\theta} ) ] =0 \ ,
\end{equation}
where we have used the fact that traceless condition of the energy
momentum tensor in 2-dimensional scalar field. From this result, we obtain
\begin{equation}
  (r \bar{g} )_{,r} = (n-3) g - (n-4) \bar{g} \ .
\end{equation}
And the equation of motion of the scalar field $\Delta \phi =0$ gives
\begin{equation}
  [ \bar{g} r^{n-2} \phi_{,r} ]_{,r}
 = -(n-2) r^{n-3} \phi_{,v} - 2 r^{n-2} \phi_{,rv} \ .
\end{equation}
The remaining equations can be derived from the above equations.
These equations motion are still difficult to solve,
so we further restrict the model to be self-similar.
More precisely, we impose the condition
\begin{equation}
   L_{\xi} g_{\mu\nu} = 2 g_{\mu\nu} \ ,
\end{equation}
where $\xi = r \partial_r + v \partial_v$.
Now, define $x=r/v$, then we obtain the function forms as
\begin{equation}
  g= g(x),\;\; \bar{g} = \bar{g} (x),\;\;
   \phi = \int^x_0 {\gamma (\eta) \over \eta} d\eta.
\end{equation}
In general, the scalar field has more freedom,
however, we do not consider it in this paper.
Indeed, that case has not do with the critical behavior
we have mentioned in sec.2. 
Using the above results, equations of motion can be written as
\begin{eqnarray}
  x g^{\prime} &=& {8\pi \over n-2} g \gamma^2  \ ,    \\
  \bar{g}^{\prime} &=& {n-3\over x} (g-\bar{g}) \ ,   \\
  g-\bar{g} &=& -{8\pi \over (n-2)(n-3)}(\bar{g} -2x)\gamma^2 \ , \\
  x(\bar{g} -2x) \gamma^{\prime} &=& - \gamma
        { (n-3) g -(n-2) x } \ ,
\end{eqnarray}
where a prime denotes the derivative with respect to $x$.
Following Brady~\cite{brady2}, we define
\begin{equation}
  y = {\bar{g} \over g} \ ,\;\; z={x\over \bar{g}} \ ,\;\;
  \xi = - \log x \ .
\end{equation}
Here we have taken into account the fact that when $v$ varies from
$0$ to $\infty$, $x$ varies from $\infty$ to $0$ and hence
$\xi$ varies from $-\infty$ to $\infty$.
Thus, the basic equations are deduced as
\begin{eqnarray}
 \dot{y} &=& -(n-3) + [{8\pi \over n-2}\gamma^2
     + (n-3) ] y   \ , \\
 \dot{z} &=& - z[n-2 - {n-3 \over y}  ]  \ , \\
 (1-2z) \dot{\gamma} &=& \gamma [{n-3 \over y} - (n-2) z ]  \ , \\
  \gamma &=&  \pm \sqrt{(n-2)(n-3)(1-y^{-1})\over 
     8\pi  (1-2z) }   \ , 
\end{eqnarray}
where a dot denotes the derivative with respect to $\xi$.
Due to the existence of the constraint equation,
the apparently 3-dimensional system reduces
to 2-dimensional dynamical system.

It is easy to find the fixed point of this system
\begin{equation}
  y_f = {n-3\over n-2} \ ,\;\; z_f =1 \ ,\;\;
  \gamma_f = \pm \sqrt{ n-2 \over 8\pi} \ .  
\end{equation}
In Fig.1, we have shown the phase flow in the 4-dimensional case. 
It certainly shows the fixed point is a saddle point.
And the apparent horizon is given by 
\begin{equation}
  g^{\mu\nu} \partial_\mu r \partial_\nu r  ={\bar{g} \over g} =y=0 \ .
\end{equation}
Then, at the fixed point, the phase flow bifurcate to the black hole
and the flat spacetime.
The line which separate both type might be called as choptuon.
As for the comparison, 
the result of 6-dimensional case is also presented in Fig.2.
It shows almost the same behavior with that of 4-dimensions.

\section{Universality of the critical phenomena}

 As the results of the previous section indicate,
the phase flow  structure is determined by the fixed point.
The important point here is that the fixed point is the saddle point,
hence the relevant mode explained below is unique.
This explains the universality of the  critical behavior
in black hole formation.
It should be noted here the similarity of our results
with those of Koike et. al.~\cite{koike} and Maison~\cite{maison}.
However, in our case as opposed to theirs,
the global phase structure is known.
So explanation is complete.
In a sense, the model considered in this paper
is the minisuperspace model for critical phenomena. 

To make the concept clear,
it is convenient to define several important operator.
The time-evolution operator is defined as
\begin{equation}
   R_{\xi} X(0) \equiv  X(\xi) \ ,
\end{equation}
where $X= (y,z)$.
Then the fixed point is characterized by
\begin{equation}
     R_{\xi} X_f = X_f   \ .
\end{equation}
 The tangent map of $R_{\xi}$ at a fixed point $X_f $ is defined as
\begin{equation}
  T_{\xi} F  \equiv  
     \lim_{\epsilon \rightarrow 0}
       { R_{\xi} (X_f + \epsilon F) - X_f \over \epsilon } \ .
\end{equation}
An eigenmode $F_i$ of $T_{\xi}$ can be defined as
\begin{equation}
  T_{\xi} F_i  = e^{\kappa_i \xi} F_i \ .
\end{equation}
A mode with $\kappa >0$ is called as a relevant mode from an apparent reason.
In the case of $\kappa =0$, we say it is marginal and
if $\kappa<0$ we say irrelevant.
Now what we have shown in the previous section
is the existence of the saddle fixed point and the critical line
on which the every point goes to the fixed point.
Hence there is a one relevant mode as is mentioned.

First of all, we must prepare the initial data, $I$,
characterized by the one parameter $p$.
A one parameter family of initial data $I$ will in general
intersect with the critical line,
and the intersection point, $X_c$ will be driven to the fixed point
under the time evolution:
\begin{equation}
  \lim_{\xi \rightarrow \infty} \mid X_c(\xi) - X_f  \mid  =0 \ . 
\end{equation}
Now, we can describe why the universality exists
in black hole formation (see Fig.3).
The intersection point is nothing but the critical point,
corresponding to the parameter $p^*$.
So the critical line corresponds to the critical solution.
If we start from the point which is different from the critical point
but close to it, it will approach to the fixed point
along the critical line.
However, at near the fixed point, it will eventually be driven away
along the direction corresponding to the relevant mode.
If the initial parameter is larger than the critical value $p^*$,
it goes to the black hole and otherwise it goes to the flat spacetime.
Apparently this behavior is determined only by the behavior
around the fixed point,
hence it is independent on the choice of the initial parameter. 

Using this picture,
we can give the general formula for the critical exponents
in n-dimensional spacetime.
Let $X_{\rm init}$ be the initial data such that
\begin{equation}
    X_{\rm init} = X_c - \epsilon F  \ ,
\end{equation}
where $F$ is the vector which deviate the initial data from
the critical value, and $\epsilon=\mid p-p^* \mid$.  
What we are interested in is the time-evolved value of this initial data,
\begin{eqnarray}
  R_{\xi} X_{\rm init} &=& R_{\xi} (X_c -\epsilon F)  \\
                      &=& X_c (\xi) - \epsilon T_{\xi} F + O(\epsilon^2) \\
                       &=& X_f - \epsilon e^{\kappa \xi} F_{\rm rel}  \ ,
\end{eqnarray}
where we used eqs. (51) and (53) in the above manipulation. 
In the final line, the relevant mode $F_{\rm rel}$ becomes dominant. 
Thus we have obtained the asymptotic behavior
of the time evolved initial data.
Let us recall the apparent horizon is given by $y=0$,
hence
\begin{eqnarray}
  y_f  &=&  \epsilon e^{\kappa \xi} F_{\rm rel,y}  \\
       &=&  \epsilon ( {v_{A.H.} \over r_{A.H.}})^{\kappa} F_{\rm rel,y} \ ,
\end{eqnarray}
where we have used the relation $ \xi = - \log {r/v}$.
Here, $F_{\rm rel,y}$ denotes the $y$-component
of the relevant mode $F_{\rm rel}$. 
Therefore, we obtain the scaling relation
\begin{equation}
    r_{A.H.} \sim \epsilon^{1\over \kappa} \ .
\end{equation}
{}From the dimensional consideration, the black hole mass scales as
\begin{equation}
  M_{B.H.} \sim  r_{A.H.}^{n-3}  \sim \epsilon^{n-3 \over \kappa} 
           \sim \mid  p-p^* \mid ^\beta  \ ,
\end{equation}
where we have identified as $ \beta = (n-3)/\kappa$.
The remaining task is the calculation of $\kappa$
by using the linear perturbation.
By substituting the linear expansion around the fixed point
\begin{equation}
 y= y_f + \delta y \ , z=z_f + \delta z \ , 
   \gamma^2 = \gamma_f^2  + \delta \gamma^2 \ , 
\end{equation}
into eqs.(43), (44) and (46), we obtain
\begin{eqnarray}
    \delta \dot y &=&  2 {n-3 \over n-2} \delta z   \ ,\\  
    \delta \dot z &=&  {(n-2)^2 \over n-3} \delta y   \ ,\\
    \delta \gamma^2 &=&  - {(n-2)^3 \over 8\pi (n-3)} \delta y
             - { 2(n-2) \over 8\pi } \delta z   \ .
\end{eqnarray}
It turns out that the fixed point is saddle point from these equations. 
It is also straightforward to calculate the eigenvalue
of the relevant mode as
\begin{equation}
  \kappa = \sqrt{ 2(n-2)} \ .
\end{equation}
Finally, we obtain 
\begin{equation}
  \beta = {n-3 \over \kappa } = { n-3 \over \sqrt{2(n-2)}} \ .
\end{equation}
Thus, we have completed our task.

\section{Conclusion}

We have investigated the higher dimensional spherically symmetric
self-similar scalar field collapse from the point of view
of the critical behavior.
In the course of the analysis,
we found the completely satisfactory explanation
of the universality of the critical behavior in black hole formation
within this restricted context.
Namely, the universality is nothing but the reflection
of the phase flow structure of the system.
In the system considered in this paper,
there is a unique fixed point which has only one relevant mode.
Then, the asymptotic behavior of the system is determined
by the behavior around the fixed point.
Hence, the critical behavior is independent on the choice
of the parameter $p$. 
This result leads us to the formula for the critical exponents
in higher dimensions.
{}From our formula, it is easy to calculate the critical exponent
as in the Table.1.
In the 4-dimensional case,
it completely agrees with the result obtained by using the exact solution. 
\begin{table}[t]
  \begin{center}
  \begin{tabular}[t]{|c|l|}
\hline
dimension $n$ & exponent $\beta$\\\hline\hline
4 & 0.5 \\
5 & 0.81650 \\
6 & 1.06066 \\
7 & 1.26491 \\
8 & 1.44338 \\
9 & 1.60357 \\
10 & 1.75 \\
11 & 1.88561 \\
12 & 2.01246 \\\hline
  \end{tabular}
  \caption{This table shows critical exponents for several dimensions.
In the 4-dimensional case, the exponent ($0.5$) completely agrees
with the result obtained by using the exact solution.
In higher dimension, the exponent takes larger value.
Especially $n \ge 6$, the exponent is larger than $1$, this fact
shows that black hole mass is smoothly connected to zero
when $p \rightarrow p^{*}$.}
  \label{tab:1}
  \end{center}
\end{table}

It is interesting to consider the relation between our work
and that of Tomimatsu~\cite{T} concerning with the quantization
of the self-similar spacetime.
In the future, we would like to attack to this problem.

\section*{Acknowledgments}

We would like to thank Werner Israel and Don Page 
for their encouragements.
We also want to say thanks to Yoshi Fujiwara and Takeshi Chiba
for their stimulating discussions on critical phenomena.
One of the authors (J.S.) thanks
the hospitality at Theoretical Physics Institute,
University of Alberta, especially David Salopek.
He also  acknowledges financial support from the Ministry of Education.


\newpage
\pagestyle{nopage} 
{\Large Figure captions}\\
\vskip .1cm
{\bf Fig.1:}
Phase flow in the 4-dimensional case is presented.
Fixed point of the 4-dimensional case is at $y=0.5$, $z=1$.
On $z=2$, initial data are set.
And there, $y$s play the role of initial parameter.
The flow curves which start from $y$ smaller than critical $y$
bifurcate to the black hole,
the others bifurcate to the flat spacetime.
\vskip .5cm
{\bf Fig.2:}
Phase flow in the 6-dimensional case is presented.
Fixed point of this case is $y=0.75$, $z=1$.
Other features are not different from the 4-dimensional case.
\vskip .5cm
{\bf Fig.3:}
Schematic figure of phase flow is shown.
Solid curves shows the phase flow, and start from $I$
which is a one parameter family of initial data.
They go along critical line and bifurcate at fixed point $X_f$.
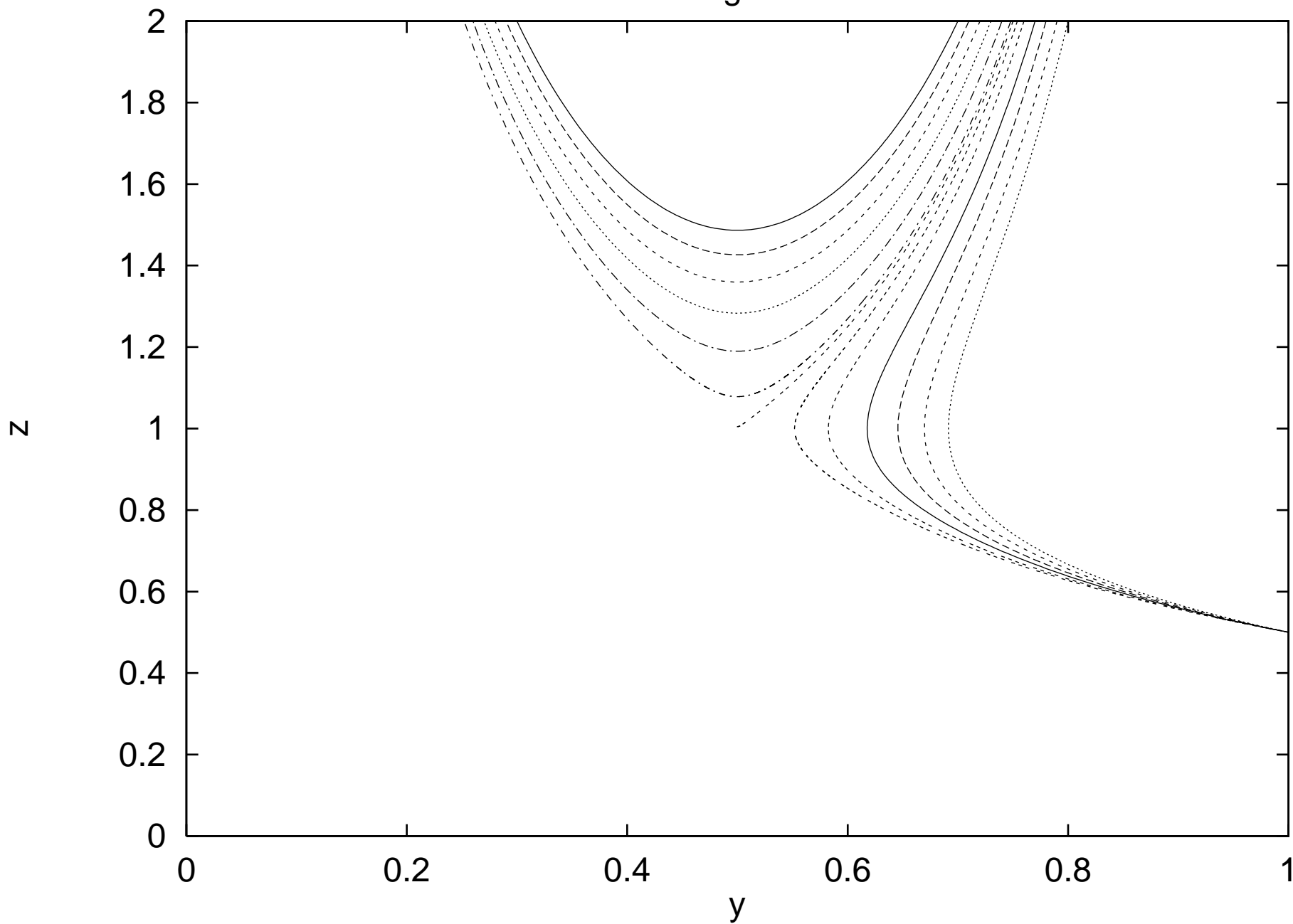
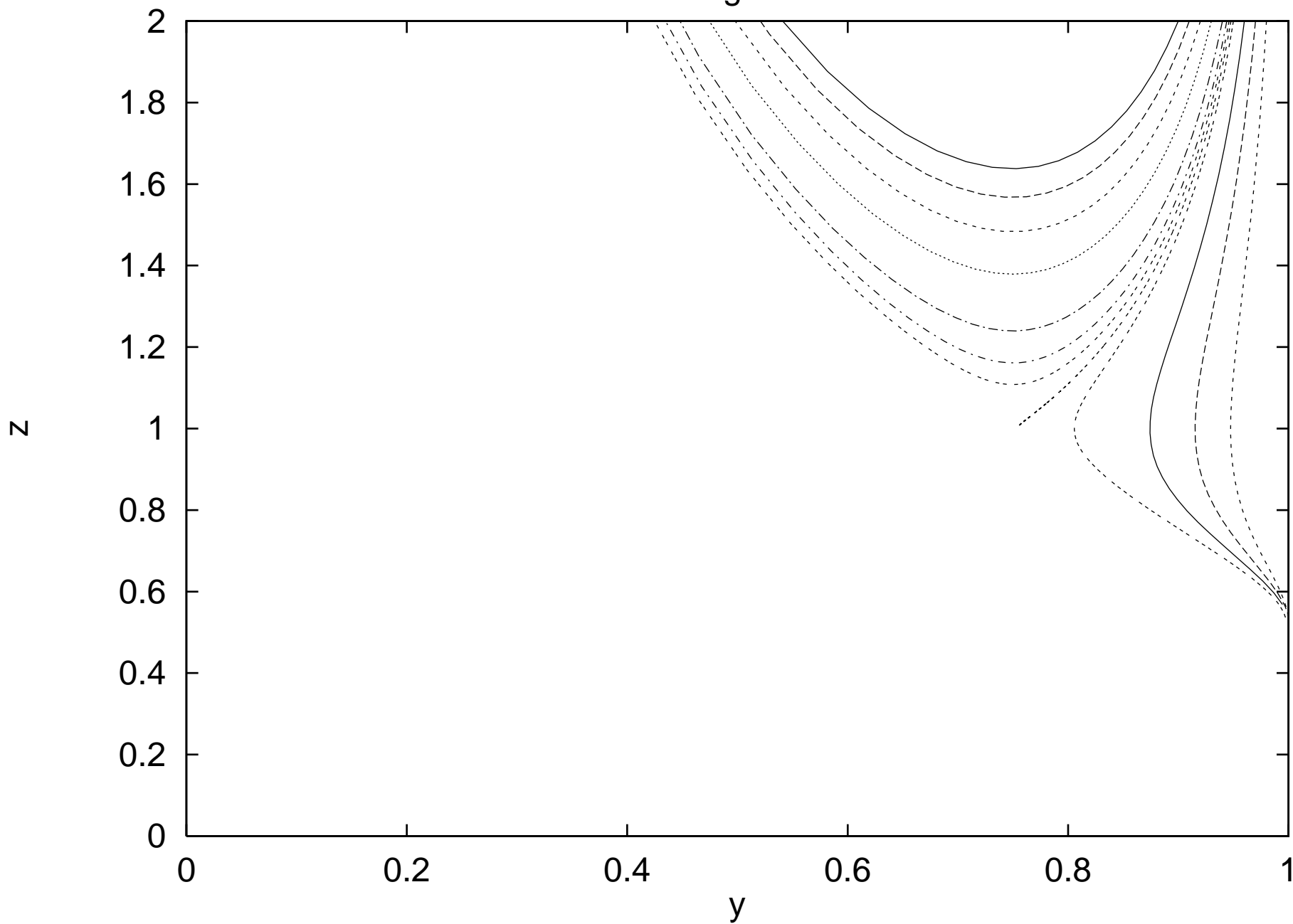
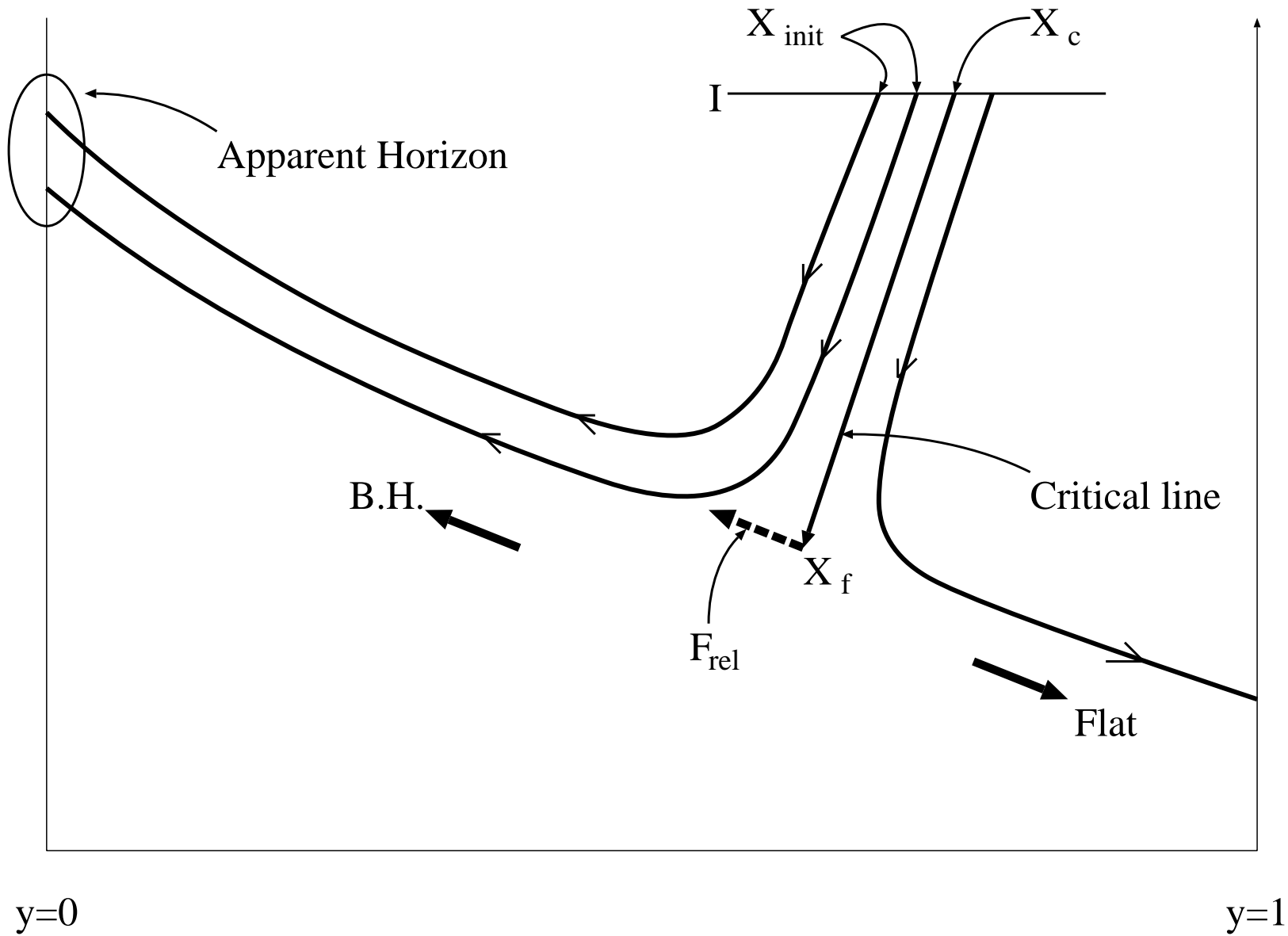
\end{document}